\documentclass[aps,prl,twocolumn,superscriptaddress,showpacs]{revtex4-1}
\usepackage{amsmath, amsthm, amsfonts, amssymb, nccmath}

\usepackage{etex}
\usepackage{epsfig}
\usepackage{dcolumn}
\usepackage{bm}
\usepackage{color}
\usepackage{lineno}

\usepackage{float}

\usepackage[normalem]{ulem}
\newcommand\redsout{\bgroup\markoverwith{\textcolor{red}{\rule[0.5ex]{2pt}{1.4pt}}}\ULon}

\newcommand{\mbc}{M_{\text{BC}}}
\usepackage{gensymb}

\usepackage[dvipsnames]{xcolor}
\usepackage{hyperref}
\hypersetup{
 pdftitle={},%
 pdfauthor={},%
 pdfsubject={},%
 pdfkeywords={},%
 pdfstartview={},%
 bookmarksopen=true, breaklinks=true, debug=true,
 colorlinks=true, linkcolor=blue, citecolor=red, urlcolor=RoyalBlue,
 hyperfigures=true
}
\def\equationautorefname~#1\null{(#1)\null}

\newcolumntype{L}[1]{>{\raggedright\let\newline\\\arraybackslash\hspace{0pt}}m{#1}}
\newcolumntype{C}[1]{>{\centering\let\newline\\\arraybackslash\hspace{0pt}}m{#1}}
\newcolumntype{R}[1]{>{\raggedleft\let\newline\\\arraybackslash\hspace{0pt}}m{#1}}

\def \b{{\cal B}}
\def \L{{\cal L}}

\begin{document}

\title{\quad\\[1.0cm] Observation of the Leptonic Decay $D^+ \to \tau^+ \nu_\tau$}

%-------- INSERT HERE ------------
\author{
\begin{small}
\begin{center}
M.~Ablikim$^{1}$, M.~N.~Achasov$^{10,d}$, P.~Adlarson$^{59}$, S. ~Ahmed$^{15}$, M.~Albrecht$^{4}$, M.~Alekseev$^{58A,58C}$, A.~Amoroso$^{58A,58C}$, F.~F.~An$^{1}$, Q.~An$^{55,43}$, Y.~Bai$^{42}$
% ...
\\
\vspace{0.2cm}
(BESIII Collaboration)\\
\vspace{0.2cm} {\it
$^{1}$ Institute of High Energy Physics, Beijing 100049, People's Republic of China\\
% ...
$^{4}$ Bochum Ruhr-University, D-44780 Bochum, Germany\\
% ...
$^{10}$ G.I. Budker Institute of Nuclear Physics SB RAS (BINP), Novosibirsk 630090, Russia\\
% ...
$^{15}$ Helmholtz Institute Mainz, Johann-Joachim-Becher-Weg 45, D-55099 Mainz, Germany\\
% ...
$^{42}$ Southeast University, Nanjing 211100, People's Republic of China\\
$^{43}$ State Key Laboratory of Particle Detection and Electronics, Beijing 100049, Hefei 230026, People's Republic of China\\
% ...
$^{55}$ University of Science and Technology of China, Hefei 230026, People's Republic of China\\
% ...
$^{58}$ (A)University of Turin, I-10125, Turin, Italy; (B)University of Eastern Piedmont, I-15121, Alessandria, Italy; (C)INFN, I-10125, Turin, Italy\\
$^{59}$ Uppsala University, Box 516, SE-75120 Uppsala, Sweden\\
% ...
$^{d}$ Also at the Novosibirsk State University, Novosibirsk, 630090, Russia\\
% ...
}\end{center}
\vspace{0.4cm}
\end{small}
\vspace{0.4cm}
}

%-------- END INSERT ------------

% -- Abstract --

\begin{abstract}
    We report the first observation of $D^+\to\tau^+\nu_\tau$ with
  a significance of $5.1\sigma$.
We measure $\b(D^+\to\tau^+\nu_\tau) =
(1.20\pm0.24_{\text{stat.}}\pm0.12_{\text{syst.}})\times10^{-3}$.
Taking the world average $\b(D^+\to\mu^+\nu_\mu) =
(3.74\pm0.17)\times10^{-4}$,
we obtain $R_{\tau/\mu} =\Gamma(D^+\to\tau^+\nu_\tau)/\Gamma(D^+\to\mu^+\nu_\mu) =
3.21\pm0.64_{\text{stat.}}\pm0.43_{\text{syst.}}$, which is consistent
with the standard model expectation of
lepton flavor universality.
Using external inputs, our results give values for the $D^+$ decay constant $f_{D^+}$
  and the Cabibbo-Kobayashi-Maskawa matrix
element $|V_{cd}|$ that are consistent with, but less precise than, other determinations.
\end{abstract}

\maketitle

%%%%%%%%%%%%%%%%%% INTRODUCTION

In the purely leptonic decay of
the charmed meson $D^+$,
the $c$ and
$\bar{d}$ quarks annihilate into
a pair of charged and neutral leptons
via a virtual $W$ boson.
(Unless otherwise noted,
charge conjugate modes are implied throughout this Letter.)
To the lowest order, the decay rate for
$D^+\to\ell^+\nu_\ell$ is given in a very simple form:
\begin{fleqn}
  \begin{equation}\label{eq:leprate}
    \setlength{\arraycolsep}{3pt}
    \text{\fontsize{9.2pt}{9.2pt}\selectfont $\Gamma(D^+\to\ell^+\nu_\ell) = \frac{G_F^2}{8\pi}f_{D^+}^2|V_{cd}|^2m_\ell^2M_{D^+}\left(1-\frac{m_\ell^2}{M_{D^+}^2}\right)^2$},
  \end{equation}
  \end{fleqn}
  \noindent where the $D^+$ mass $M_{D^+}$, the masses
  of the charged leptons  $m_\ell$  ($\ell = e^+$, $\mu^+$, or
  $\tau^+$), and the Fermi coupling constant $G_F$
  are  known to great precision~\cite{pdg2018}.
  Because of this, measuring $\b(D^+\to\ell^+\nu_\ell)$ ($\b_{\ell\nu}$) allows
  determination of the product $f_{D^+}^2|V_{cd}|^2$
of the $D^+$ decay constant and the square of the $c\to d$
Cabibbo-Kobayashi-Maskawa (CKM) matrix element.
One can then extract $|V_{cd}|$ by using the predicted value of
$f_{D^+}$, e.g.,
from lattice quantum chromodynamics (LQCD),
or obtain  $f_{D^+}$ by using the experimentally measured  $|V_{cd}|$
to test the LQCD prediction.
Such studies have been done
using the muonic mode
$D^+\to\mu^+\nu_\mu$~(\cite{bes3munu},\cite{cleoDpmunu}),
which is a simple two-body decay
with a clear experimental signature.
The energetic track produced in this decay can be reconstructed very efficiently with minimal systematic uncertainty.

Experimental information about $D^+\to\tau^+\nu_\tau$ 
is more sparse, with only an upper limit of
$1.2\times10^{-3}$ on $\b_{\tau\nu}$ at a $90\%$ confidence level (C.L.)~\cite{pdg2018}
that was set by the CLEO Collaboration~\cite{cleoDpmunu}.
Measuring $\b_{\tau\nu} $
is an important check of the standard model, which predicts the ratio
of the $\tau^+\nu_\tau$ and $\mu^+\nu_\mu$ decay rates.
Applying Eq.~(\ref{eq:leprate}) to both $D^+\to\mu^+\nu_\mu$ and 
$D^+\to\tau^+\nu_\tau$, we find
\begin{equation}\label{eq:R}
 R_{\tau/\mu} =
 \frac{\Gamma(D^+\to\tau^+\nu_\tau)}{\Gamma(D^+\to\mu^+\nu_\mu)} =
 \frac{m^2_{\tau}(1-\frac{m_{\tau}^2}{M_{D^+}^2})^2}{m^2_\mu(1-\frac{m_{\mu}^2}{M_{D^+}^2})^2}=2.67,
\end{equation}
which provides
a clean
test of the standard model
expectation of lepton flavor universality.
Deviation from the expected value of $R_{\tau/\mu}$
could signify contributions of a charged intermediate boson
that couples to the leptons differently, e.g., through
a leptoquark~\cite{leptoquark}.
The fact that $\b_{\tau\nu}$
has not been measured previously,
  together with the
recent hints of possible violation of lepton
universality
  in $B$ decays~\cite{LUV},
  establishes that $R_{\tau/\mu}$ is an
important quantity to determine experimentally.
We note, however, that in some supersymmetric models, such as the 
two-Higgs-doublet model~\cite{chagedhiggs},
the charged Higgs couples to the lepton mass leading to a mass dependence 
identical to that from the $W$ boson process, including its helicity suppression.  
Thus, Eq.~(\ref{eq:leprate}) is modified by a factor that does not depend on 
the lepton masses, leaving $R_{\tau/\mu}$ unchanged.

From the standard model prediction $R_{\tau/\mu} = 2.67$ and $\b_{\mu\nu} = (3.74\pm0.17)\times10^{-4}$~\cite{pdg2018},
one expects  $\b_{\tau\nu} = (1.01\pm0.05)\times10^{-3}$, which is very close to CLEO's
upper limit based on
$818$ pb$^{-1}$ of $e^+e^-$ annihilation data.
In this Letter, we report the first observation of  $D^+\to\tau^+\nu_\tau$
and the measurement of its branching fraction with an $e^+e^-$
annihilation sample produced at the Beijing Electron Positron Collider
(BEPCII)~\cite{bepcii}
near the nominal mass of the $\psi(3770)$ resonance,
$\sqrt{s} = 3.773$~GeV, with
an integrated luminosity of $2931.8$~pb$^{-1}$~\cite{bes3lumi} 
collected with the BESIII detector.

%%%%%%%%%%%%%%%%%% BESIII DETECTOR and MC sample

BESIII is a cylindrical detector with a solid angle
coverage of $93\%$ of $4\pi$.
The detector consists of a Helium-gas-based main drift chamber (MDC),
a plastic scintillator time-of-flight system, a CsI(Tl)
electromagnetic calorimeter (EMC), a superconducting solenoid
providing a $1.0$ T magnetic field, and muon counters.
The charged particle momentum resolution is $0.5\%$ at a transverse
momentum of $1$~GeV$/c$.
The photon energy resolution at $1$~GeV is $2.5\%$ in the central
barrel region and
$5.0\%$
in the end cap region.
More details about the design and performance of BESIII are
given in Ref.~\cite{bes3detector}.

Detection efficiencies and background processes are
determined with a
Monte Carlo (MC) simulation sample
with an equivalent luminosity
roughly $10$ times larger than the data set.
It consists of events from $e^+e^-\to \psi(3770)\to D\bar{D}$,
$e^+e^-\to q\bar{q}$ ($q = u, d, s)$, $e^+e^-\to\gamma J/\psi$,
$e^+e^-\to\gamma\psi(3686)$, and $e^+e^-\to\tau^+\tau^-$.
The effects of
initial- and final-state radiation
are simulated by the {\sc KKMC} generator~\cite{kkmc}
and the {\footnotesize PHOTOS} package~\cite{photos}, respectively.
The generated four-momenta are propagated into {\sc EvtGen}~\cite{evtgen}, which
simulates decays using known rates~\cite{pdg} and correct
angular distributions.
We generate charmonium decays not accounted for by
exclusive measurements with {\sc Lundcharm}~\cite{lundcharm}.
Finally, the detector response is simulated with
{\sc geant4}~\cite{geant}.

%%%%%%%%%%%%%%%%%% Analysis Procedure

We measure $\b_{\tau\nu} $ by reconstructing 
$\tau^+$ via $\tau^+\to\pi^+\bar{\nu}_{\tau}$, which
has the feature of only a single charged track
from the $D^+$ decay.
Because pions and muons are charged particles with
similar masses, the BESIII selection of pion tracks
based on specific-ionization and time-of-flight
measurements also accepts muon tracks with comparable
efficiency ($>90\%$),
allowing simultaneous measurement of $\b_{\tau\nu} $ and
$\b_{\mu\nu}$.
For this analysis our main result is obtained by fixing
$\b_{\mu\nu}$ to the
world average of $(3.74\pm0.17)\times10^{-4}$~\cite{pdg2018}
to maximize our statistical sensitivity for measuring $\b_{\tau\nu}$.
We also perform a cross-check of our result by
measuring $\b_{\mu\nu}$ and $\b_{\tau\nu}$ simultaneously.

This analysis employs a double-tag technique, pioneered by the Mark
III Collaboration~\cite{mark3}.
We obtain the branching fraction by reconstructing
$D^+\to\tau^+(\to\pi^+\bar{\nu}_{\tau})\nu_\tau$ in events with
$D^-$ decays reconstructed in one of the six tag modes listed in Table~\ref{tab:stinfo}:
\begin{equation}\label{eq:DT}
 \b_{\tau\nu} = \frac{N_{\tau\nu}}{\sum_{i}{N_{\text{tag}}^{i}(\epsilon^i_{\tau\nu}/\epsilon^i_{\text{tag}})}}. \\
\end{equation}
\noindent In Eq.~(\ref{eq:DT}) $N_{\tau\nu}$ is the number of events with any $D^-$ tag and a 
$D^+\to\tau^+(\to\pi^+\bar{\nu}_\tau)\nu_\tau$ candidate, $\epsilon^i_{\tau\nu}$ is the
signal selection efficiency including $\b(\tau^+\to\pi^+\bar{\nu}_\tau)$
for an event with a $D^-$ in the $i\text{th}$ tag mode, and 
$N_{\text{tag}}^{i}$ and $\epsilon^i_{\text{tag}}$ are the number of
tag and reconstruction
efficiency for $D^-$ tags in mode $i$.

\begin{figure}[htbp]
\centering
 \includegraphics[keepaspectratio=true,width=3.1in,angle=0]{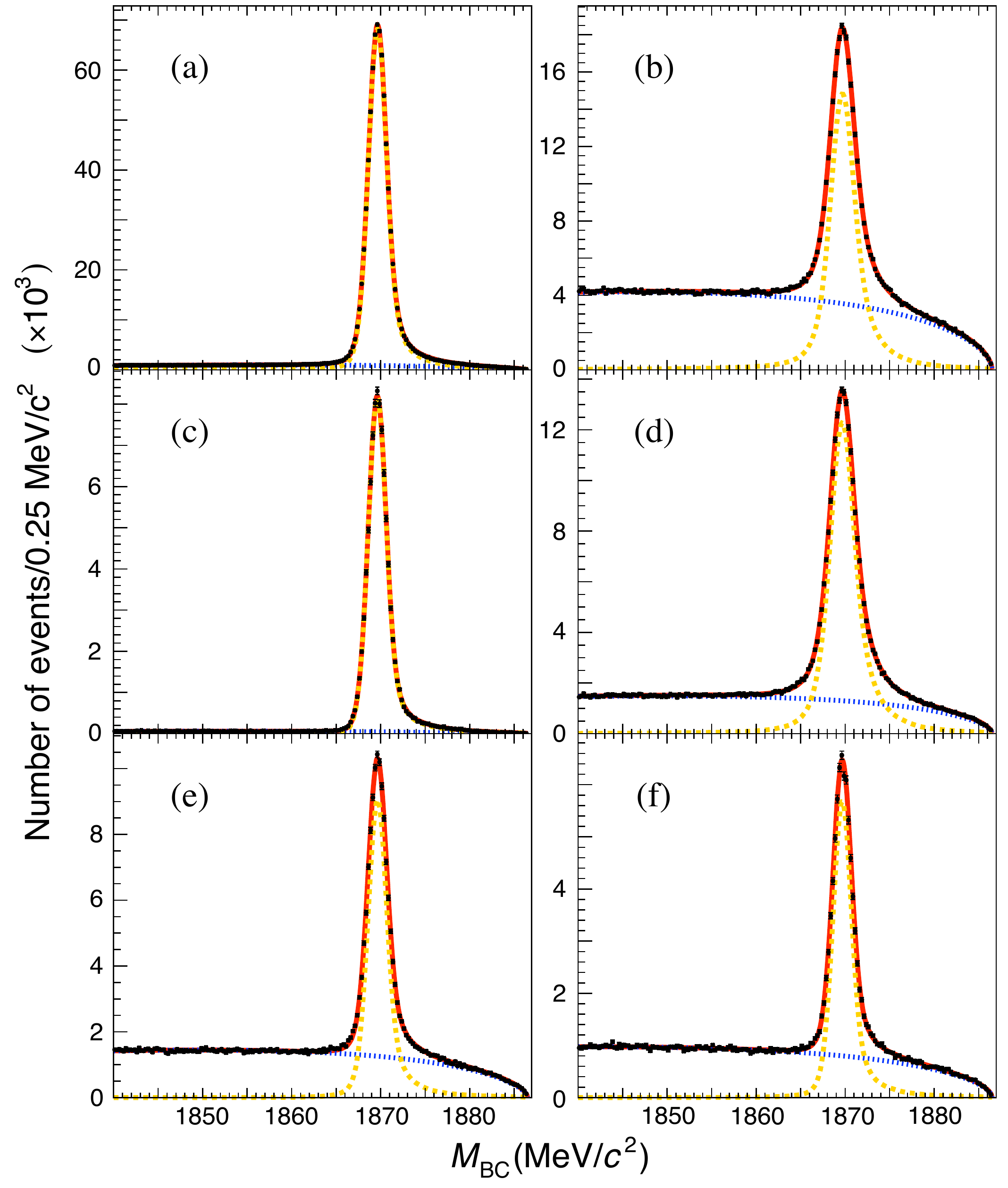}
 \caption{Fits to $\mbc$ distributions of single-tag $D^-$ candidates for the full data
  sample for tag modes
$D^-\to$  (a) $K^+\pi^-\pi^-$,  (b) $K^+\pi^-\pi^-\pi^0$,
(c) $K_S^0\pi^-$, (d) $K_S^0\pi^-\pi^0$, (e) $K_S^0\pi^-\pi^-\pi^+$,
and (f) $K^+K^-\pi^-$.
Red lines are the overall fits, while the yellow-dashed (blue-dotted) lines
are the fitted signals (backgrounds).}
\label{fig:ST}
\end{figure}

%%%%%%%%%%%%%%%%%% ST
In selecting tags
our criteria for the final-state particles
are identical to those in Ref.~\cite{boss664}.
In each event, we allow only one $D$ candidate for
  a given
tag mode separately for $D^+$ and $D^-$,
  following the method of Ref.~\cite{dErange}.
For each tag mode, we extract $N_{\text{tag}}^{i}$
from distributions of beam-constrained mass
$\mbc c^2 = \sqrt{E^2_{\text{beam}} - |\vec{p}_{\text{tag}}c|^2}$, where
$\vec{p}_{\text{tag}}$ is the three-momentum of the tag $D^-$ candidate
and $E_{\text{beam}}$ is the beam energy in the center-of-mass system
of the $\psi(3770)$. 
We fit to these $\mbc$ distributions with MC-based signal shapes that
are convolved with a Gaussian to accommodate resolution
differences between simulation and data. The background shape
is parametrized with an ARGUS function~\cite{argus}.
Figure~\ref{fig:ST} shows the fits to $\mbc$ distributions.
To select the tag, 
we require that $1863<\mbc<1877$~MeV$/c^2$~\cite{MBCrange}.
Table~\ref{tab:stinfo} shows $N^i_{\text{tag}}$, $\epsilon^i_{\text{tag}}$, 
and $\epsilon^i_{\tau\nu}$ for all tag modes.

\begin{table}[htb]
  \caption{Single-tag efficiencies ($\epsilon^i_{\text{tag}}$)
    and yields ($N_{\text{tag}}^{i}$), and signal selection
    efficiencies ($\epsilon^i_{\tau\nu}$).
    Efficiencies are corrected for
$\b(K_S^0\to\pi^+\pi^-)$ and $\b(\pi^0\to\gamma\gamma)$.
}
\def\1#1#2#3{\multicolumn{#1}{#2}{#3}}
\label{tab:stinfo}
\begin{center}
\scalebox{1.0}
{
  \begin{tabular}{l r  c c }
    \hline\hline
    \multicolumn{1}{c}{Tag modes, $i$}
    & \multicolumn{1}{c}{$N_{\text{tag}}^{i}$ ($\times 10^3$)}
    & $\epsilon^i_{\text{tag}}$ (\%) 
    & $\epsilon^i_{\tau\nu}$ (\%)\\
\hline
$K^+\pi^-\pi^-$                &   $797.6\pm1.0$ & $51.06\pm0.03$ & $3.6\pm0.1$\\
$K^+\pi^-\pi^-\pi^0$        &  $245.1\pm0.7$ & $25.18\pm0.03$ & $2.1\pm0.1$\\
$K_S^0\pi^-$                      &  $92.6\pm0.3$ & $50.66\pm0.07$ & $4.0\pm0.1$\\
$K_S^0\pi^-\pi^0$              &  $206.3\pm0.6$ & $26.09\pm0.03$ & $2.1\pm0.1$\\
$K_S^0\pi^-\pi^-\pi^+$     &  $110.2\pm0.4$ & $26.75\pm0.04$ & $2.2\pm0.1$\\
$K^+K^-\pi^-$                    &  $68.1\pm0.3$ & $40.38\pm0.08$ & $3.1\pm0.1$\\
\hline\hline
\end{tabular}
}
\end{center}
\end{table}

%%%%%%%%%%%%%%%%%% DT
\newpage
Once we select the tag,
we require that there be only one additional charged
track and that it have
charge opposite to that of the tag.
It must originate
within $1$~cm ($10$~cm) from the beam interaction point
in the plane transverse to (along) the beam direction,
be within the fiducial region for reliable track reconstruction
($|\cos{\theta}|<0.93$, where $\theta$ is the polar angle with
  respect to the direction of the positron beam), and match a shower in the EMC.
Furthermore, to distinguish between
$\pi$-like and $\mu$-like tracks, we rely on the minimum-ionizing character
of the $\mu$ track,
which has
a mean energy deposit of
$E_{\text{EMC}}\simeq 200$~MeV, as was
done in Refs.~\cite{bes3munu, cleoDpmunu}.
Thus we partition the selected events into two samples,
one with $\mu$-like tracks ($E_{\text{EMC}}\le 300$~MeV)
and the other with $\pi$-like tracks  ($E_{\text{EMC}}>300$~MeV).
The first portion includes $99\%$ of the muon tracks
from $D^+\to\mu^+\nu_\mu$, while the second has $44\%$ of the
pion tracks from $D^+\to\tau^+\nu_\tau$, $\tau^+\to\pi^+\bar{\nu}_\tau$.

To suppress backgrounds further, we apply four
additional requirements, which are optimized based on MC calculations.
(1) $E_{\text{EMC}}/|\vec{p}c| < 0.95$ for the $\pi$-like sample, where
$\vec{p}$ is the signal track 
momentum measured by the MDC.
As this variable sharply peaks around $1$ for an electron,
this requirement
suppresses events from semileptonic decays like $D^+\to K^0_L e^+\nu_e$.
(2) $E_{\text{max}} < 300$~MeV for both samples, where
$E_{\text{max}}$ is the maximum energy of all
EMC showers that are not assigned to any
charged track or neutral EMC shower in the
reconstruction of both
  $D^+$ and $D^-$. This suppresses
events with extra particles, including misreconstructed
neutral pions. 
(3) $|\cos{\theta_{\text{missing}}}| < 0.95 (0.75)$ for
  the $\mu$ ($\pi$)-like sample, where $\theta_{\text{missing}}$ is
  the polar angle of the missing momentum
 $\vec{p}_{\text{missing}} = -\vec{p}_{D^-} - \vec{p}_{\mu(\pi)}$,
$\vec{p}_{D^-} = \hat{p}_{D^-}\sqrt{(E_{\text{beam}}/c)^2-(M_{D^+}c)^2}$,
and $\hat{p}_{D^-}$ is the unit momentum vector of the $D^-$.
  This ensures that $\vec{p}_{\text{missing}}$ points to an active region
  of the detector.
  (4) $\alpha > 25\degree (45\degree)$ for 
  the $\mu$ ($\pi$)-like sample, where $\alpha$ is the opening angle
  between $\vec{p}_{\text{missing}}$ and the direction of the
  most energetic unassigned shower.
  A shower from an asymmetric decay of $\pi^0$
  or from $K^0_L$ tends to deposit energy in the EMC in the
  $\vec{p}_{\text{missing}}$ direction.
The minimum required energy of the unassigned shower
is  $25$~MeV for $|\cos{\theta}|<0.8$ and
$50$~MeV for $0.86<|\cos{\theta}|<0.93$.

  Signals are extracted from the distributions of missing mass-squared
  $M_{\text{miss}}^2$ $= E_{\text{missing}}^2 - |\vec{p}_{\text{missing}}c|^2$,
where $E_{\text{missing}} = E_{\text{beam}} - E_{\mu(\pi)}$.
Events from $D^+\to\mu^+\nu_\mu$ peak around $M_{\text{miss}}^2$ $=0$,
and the ones from
$D^+\to\tau^+\nu_\tau$, where $\tau^+\to\pi^+\bar{\nu}_\tau$, also tend to populate near 
 $M_{\text{miss}}^2$ $=0$ because $m_\tau\simeq M_{D}$.

 We expect peaking backgrounds from $D^+\to\pi^0\pi^+$
and
$D^+\to K^0_L\pi^+$. The first is relatively small, but is
close to $M_{\text{miss}}^2$ $=0$. The latter peaks away
from $M_{\text{miss}}^2$ $=0$ at m$^2_{K^0}$, but is a concern
because of an expected rate of
$40$ times the expected signal.

\begin{figure}[htbp]
\centering
\includegraphics[keepaspectratio=true,width=3.1in,angle=0]{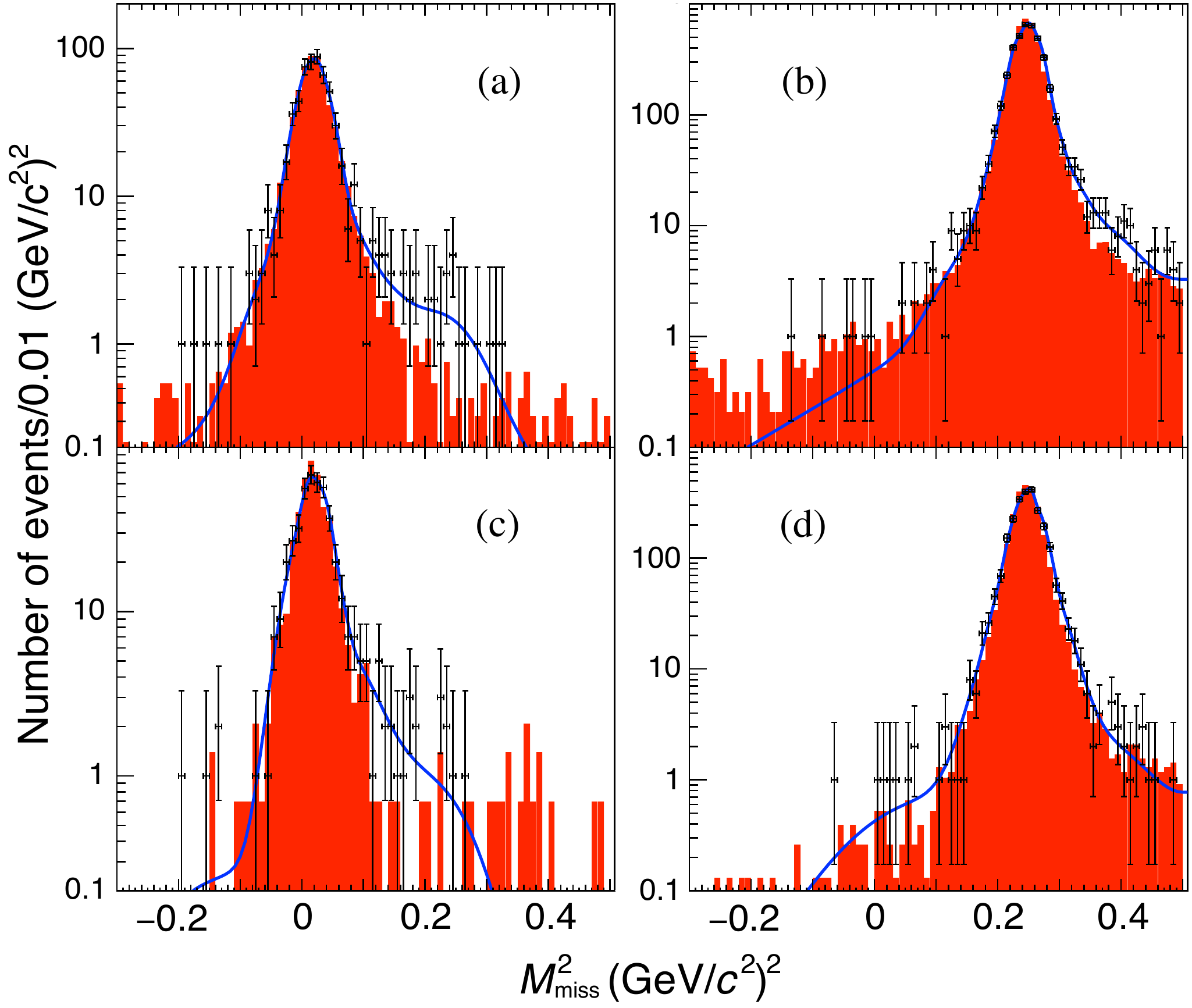}
 \caption{$M_{\text{miss}}^2$ distributions of
   $D^+\to\pi^0\pi^+$ (a), (c) and  $D^+\to K^0_S(\to\pi^+\pi^-)\pi^+$
   (b), (d) events from data (black points) for the $\mu$-like (a), (b)
   and $\pi$-like (c), (d) samples.
   The blue lines are the PDFs derived from the black points,
   while the red-shaded histograms are true
   $D^+\to\pi^0\pi^+$ and  $D^+\to K^0_L\pi^+$ MC events
   with all selection criteria applied.
 }
\label{fig:cntr}
\end{figure}

We use data-based control samples to construct
the probability density functions (PDFs)
to represent these two peaking backgrounds.
The black points in  Fig.~\ref{fig:cntr} show $M_{\text{miss}}^2$ distributions from
 exclusively reconstructed $D^+\to\pi^0(\to\gamma\gamma)\pi^+$ (left column)
 and  $D^+\to K^0_S(\to\pi^+\pi^-)\pi^+$  (right column) events
 in which we treat the $K^0_S$ and the $\pi^0$
 as missing particles, respectively.
 The red-shaded histograms are from
 true $D^+\to\pi^0\pi^+$ and $D^+\to K^0_L\pi^+$ MC events
 after applying all signal selection
 criteria, scaled to the same sizes as the data. Agreement
 between the shapes of the expected distributions and our control
 samples is excellent.
 We generate the corresponding PDFs by smoothing the distributions
 of the data points
 by the kernel estimation method~\cite{keypdf}.
\noindent Additional peaking backgrounds from
$D^+\to\eta(\to\gamma\gamma)\pi^+$ and
$D^+\to K^0_S(\to\pi^0\pi^0)\pi^+$ are also considered, but
both are small and peak away from $M_{\text{miss}}^2$ $=0$.
For these two small backgrounds, we use the MC events to predict the PDF.

\begin{figure*}[hbtp]
  \centering
  \includegraphics[keepaspectratio=true,width=6.8in,angle=0]{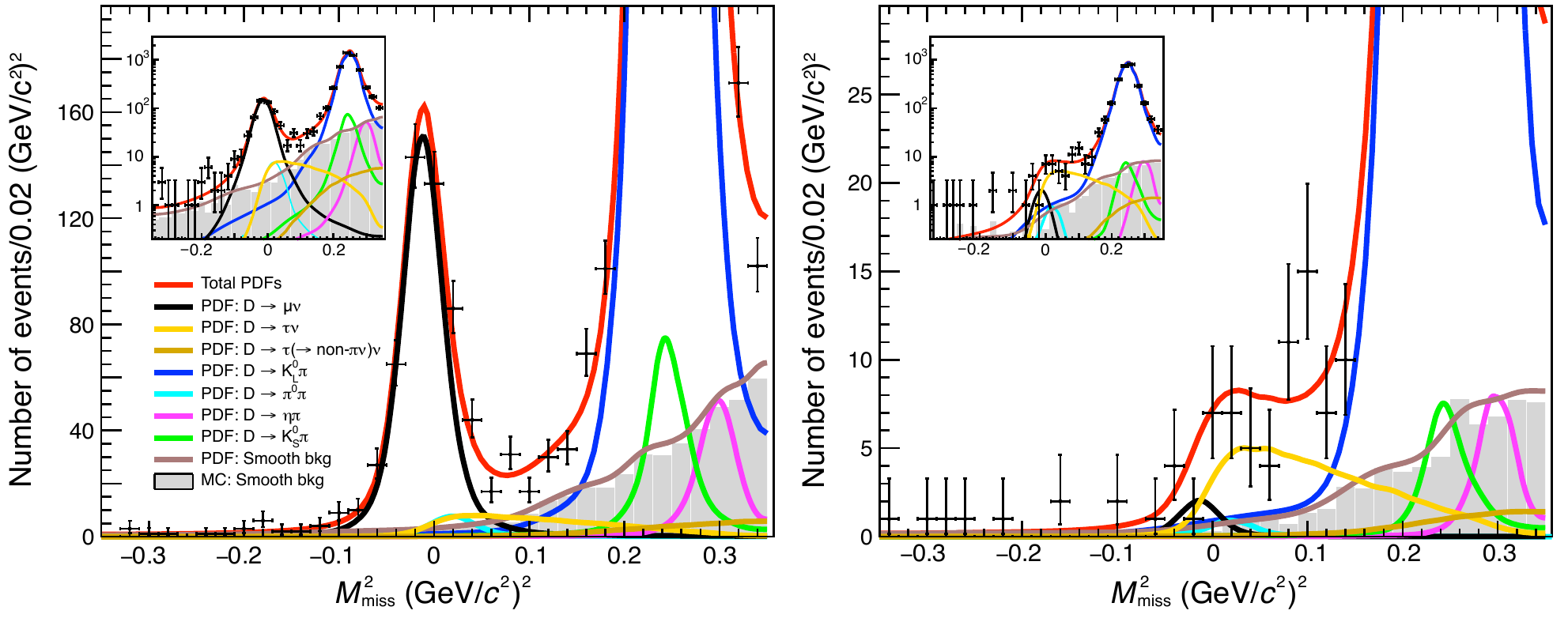}
 \caption{Fits to $M_{\text{miss}}^2$ distributions of the $\mu$-like (left)
   and $\pi$-like (right) samples.
   The black points are data and gray-shaded histograms are
   MC-predicted smooth background components scaled
   to the data size based on the known production cross sections and
   measured integrated luminosity.
   The insets show the same distributions with logarithmic scales.
 }
\label{fig:DT}
\end{figure*}

We perform an unbinned simultaneous maximum likelihood fit
to the $\mu$- and $\pi$-like samples. The signal PDFs are based on
MC events, including  $D^+\to\tau^+\nu_\tau$ with $\tau^+$ final states
other than $\pi^+\bar{\nu}_\tau$.
This contribution is
dominated by $\tau^+\to\mu^+\nu_\mu\bar{\nu}_\tau$ and $\pi^+\pi^0\bar{\nu}_\tau$
in the $\mu$-like sample, while the $\pi$-like sample mostly contains
$\tau^+\to e^+\nu_ e\bar{\nu}_\tau$ and $\pi^+\pi^0\bar{\nu}_\tau$.
To take into account the $M_{\text{miss}}^2$ resolution difference
between the data and the MC samples, the PDFs of the signal
and of the backgrounds are smeared using a Gaussian.
The Gaussian mean and width are free parameters of the fit.
The remaining background (``smooth background'') comes from
other well known $D$ decays, such as semileptonic decays,
as well as continuum events. It is represented
by the smoothed MC prediction.
We fix the sizes of $D^+$ decays to $\mu^+\nu_\mu$,
$\pi^0\pi^+$, $\eta\pi^+$, and $K^0_S\pi^+$
according to Ref.~\cite{pdg2018}, while we leave the normalizations
for decays to $\tau^+\nu_\tau$ and $K^0_L\pi^+$, as well as 
the smooth background as free fit parameters.
The ratio of the normalizations of the smooth background
between the $\mu$-like and $\pi$-like samples is constrained
based on the MC prediction.
Applying this fitting procedure to the $D\bar{D}$ MC sample,
we obtain the signal selection efficiencies $\epsilon^i_{\tau\nu}$
for each tag mode listed in Table~\ref{tab:stinfo}.

Figure~\ref{fig:DT} shows the simultaneous fit to data, which yields
$137\pm27$ signal events. This corresponds to
$\b_{\tau\nu} = (1.20\pm0.24_{\text{stat.}})\times10^{-3}$.

As a cross check, we treat the $D^+\to\mu^+\nu_\mu$ component
 as a free fit parameter and obtain
 $\b_{\mu\nu} = (3.70\pm0.20_{\text{stat.}})\times10^{-4}$,
 along with $\b_{\tau\nu} = (1.21\pm0.24_{\text{stat.}})\times10^{-3}$.
 The obtained $\b_{\mu\nu}$ 
 is consistent with both the world average of
 $(3.74\pm0.17)\times10^{-4}$~\cite{pdg2018}
 and
 the recent BESIII measurement of
 $(3.71\pm0.19_{\text{stat.}}\pm0.06_{\text{syst.}})\times10^{-4}$~\cite{bes3munu}.
 The agreement with the latter measurement provides
 independent confirmation,
 as Ref.~\cite{bes3munu} uses muon counter information
and is based on
 simulations of the signal
efficiency
and the background that are different from the current work.

 %%%%%%%%%%%%%%%%%% Systematic uncertainties
The total systematic uncertainty is dominated by two sources.
The first is the uncertainty on $\b_{\mu\nu}$,
which is fixed to the value from  Ref.~\cite{pdg2018}.
The second is the uncertainty due to the assumed shapes of the smooth background.
For this we vary the shape by changing the
$e^+e^-\to\psi(3770)\to D\bar{D}$ and  $e^+e^-\to q\bar{q}$
  cross sections from the defaults in our MC calculations.
  We also consider two different values of the smoothing
  parameter $\rho$ in the Gaussian
  kernel estimation method~\cite{keypdf}, $\rho=1$
  (the author's suggestion) and $\rho=2$.
  The dependence on the choice of $300$~MeV for the
  boundary between $\pi$- and $\mu$-like
  samples, which potentially changes the shapes of the smooth backgrounds,
  is also assessed by varying it by $\pm50$~MeV.

Other sources of systematic uncertainty
are also considered. The uncertainty in the signal track reconstruction
efficiency has been obtained from previous BESIII studies of
double-tagged
$D\bar{D}$ events.
The uncertainty in  $\b(\tau^+\to\pi^+\bar{\nu}_\tau)$ is from Ref.~\cite{pdg2018}.
Statistical uncertainties in the tag counts in data and MC
calculations are taken
directly from the respective samples.
Variations in the fit ranges, selection windows, binning, and signal
and background parametrizations are used to probe uncertainties in the
tag-side fits.
We estimate uncertainties due to the 
$E_{\text{EMC}}/|\vec{p} c|$ and $E_{\text{max}}$ criteria with
double-tagged events including $D^+\to K^0_S\pi^+$.
Uncertainties from the cuts on $|\cos{\theta_{\text{missing}}}|$
and $\alpha$ are estimated with fully reconstructed 
$D^0\to K^-e^+\nu_e$ events.
Possible mismodeling of efficiencies due to multiplicity differences among
$D$ decay modes is estimated based on a study of tracking and particle
identification efficiencies in different event environments.
The uncertainty due to the normalization of the peaking
backgrounds, and the ratio of smooth background sizes
between $\mu$- and $\pi$-like samples
in the $M_{\text{miss}}^2$ fit are estimated by studies of
the $D^+\to K^0_S\pi^+$ control sample
and by varying parametrizations and branching
fractions, respectively.
The $D^+\to\tau^+\nu_\tau$ signal-shape dependence is estimated
by altering the mixture of $\tau^+$ decay modes.
Table~\ref{tab:syst_summary} summarizes the systematic uncertainty estimate.

\begin{table}[htb]
\caption{Summary of relative systematic uncertainties in units of
  $10^{-2}$.
}
\def\1#1#2#3{\multicolumn{#1}{#2}{#3}}
\label{tab:syst_summary}
\begin{center}
\scalebox{0.96}
{
\begin{tabular}{c c }
\hline\hline
Source & $\Delta\b_{\tau\nu}$ \\
  \hline
   $\Delta\b(D^+\to\mu^+\nu_\mu)$ & $6.9$ \\
  Shape of smooth background & $4.2$\\
  $\pi^+$ tracking & $1.0$ \\
  $\Delta \b(\tau^+\to\pi^+\bar{\nu}_\tau)$ & $0.5$\\
  Stat. uncertainties from tag side and MC calculations & $2.2$ \\
  Fitting scheme on tag side & $0.5$  \\
  Requirement on  $E_{\text{EMC}}/|\vec{p}c|$ & $2.5$ \\
  Requirement on $E_{\text{max}}$ & $2.2$  \\
  Requirements on $|\cos{\theta_{\text{missing}}}|$ and $\alpha$ & $2.1$ \\
  Tag bias & $0.1$ \\
  Normalizations of small peaking backgrounds & $1.9$ \\
  Relative size of smooth background components & $2.5$\\
  Signal shape of $D^+\to\tau^+\nu_\tau$ & $1.1$\\
\hline
Total systematic uncertainty & $9.9$ \\
\hline\hline
\end{tabular}
}
\end{center}
\end{table}

%%%%%%%%%%%%%%%%%% Conclusions

Using the $2.93$~fb$^{-1}$ data sample taken at
$\sqrt{s} = 3.773$~GeV,
we measure
$\b_{\tau\nu} = (1.20\pm0.24_{\text{stat.}}\pm0.12_{\text{syst.}})\times10^{-3}$
using $\b_{\mu\nu} = (3.74\pm0.17)\times10^{-4}$.
 The signal significance including the systematic
 uncertainty is $5.1\sigma$, calculated via
   $\sqrt{-2\times\ln{\L_{\text{null}}/\L}}$,
 where $\L_{\text{null}}$ and $\L$ are
   likelihood values without and
 with $D^+\to\tau^+\nu_\tau$, respectively.
 This is the first measurement of the branching
 fraction of $D^+\to\tau^+\nu_\tau$ to date.
 With $\b_{\mu\nu} = (3.74\pm0.17)\times10^{-4}$~\cite{pdg2018}, we find
 $R_{\tau/\mu} = 3.21\pm0.64_{\text{stat.}}\pm0.43_{\text{syst.}}$, which is consistent with
 the standard model prediction of $2.67$.
 From Eq.~(\ref{eq:leprate}), with the inputs shown in
 Table~\ref{tab:inputs}
 and assuming $|V_{cd}| = 0.22438\pm0.00044$ from the global
 fit~\cite{pdg2018}, we obtain
 \[
   f_{D^+} = 224.5\pm22.8_{\text{stat.}}\pm11.3_{\text{syst.}}\pm0.9_{\text{ext-syst.}}~\text{MeV},
 \]
 \noindent where the last uncertainty is due to external input parameters.
 This is consistent with the average
 between the recent four-flavor LQCD
 predictions of Refs.~\cite{latticeetm, latticefnal},
$f_{D^+} = 212.6\pm0.6$~MeV,
as well as with the
experimental results for
$D^+\to\mu^+\nu_\mu$ from
the BESIII~\cite{bes3munu} and the CLEO~\cite{cleoDpmunu} Collaborations.

 \begin{table}[htb]
   \caption{External input parameters with uncertainties
     from Ref.~\cite{pdg2018}.
}
\def\1#1#2#3{\multicolumn{#1}{#2}{#3}}
\label{tab:inputs}
\begin{center}
\scalebox{1.0}
{
\begin{tabular}{c l  }
  \hline\hline
  Parameter & \multicolumn{1}{c}{Value} \\
  \hline
  $m_\mu$ & $0.1056583745(24)$~GeV \\
  $m_\tau$ & $1.77686(12)$~GeV \\
  $M_{D^+}$ & $1.86965(5)$~GeV \\
  $\tau_{D^+}$ & $1.040(7)$~ps\\
 $G_F$ & $1.1663787(6)\times10^{-5}$~GeV$^{-2}$ \\
  \hline\hline
\end{tabular}
}
\end{center}
\end{table}

\noindent Taking the average prediction for
 $f_{D^+}$ from \cite{latticeetm} and  \cite{latticefnal},
we find
\[
  |V_{cd}| = 0.237\pm0.024_{\text{stat.}}\pm0.012_{\text{syst.}}\pm0.001_{\text{ex-syst}}.
\]
This is consistent with both the world average $|V_{cd}| =
0.218\pm0.004$~\cite{pdg2018} and the global fit result~\cite{pdg2018}.

The BESIII collaboration thanks the staff of BEPCII and the IHEP computing center for their strong support. This work is supported in part by National Key Basic Research Program of China under Contract No. 2015CB856700; National Natural Science Foundation of China (NSFC) under Contracts No. 11625523, No. 11635010, and No. 11735014; National Natural Science Foundation of China (NSFC) under Contract No. 11835012; the Chinese Academy of Sciences (CAS) Large-Scale Scientific Facility Program; Joint Large-Scale Scientific Facility Funds of the NSFC and CAS under Contracts No. U1532257, No. U1532258, No. U1732263, and No. U1832207; CAS Key Research Program of Frontier Sciences under Contracts No. QYZDJ-SSW-SLH003 and No. QYZDJ-SSW-SLH040; 100 Talents Program of CAS; INPAC and Shanghai Key Laboratory for Particle Physics and Cosmology; German Research Foundation DFG under Contract No. Collaborative Research Center CRC 1044, FOR 2359; Istituto Nazionale di Fisica Nucleare, Italy; Koninklijke Nederlandse Akademie van Wetenschappen (KNAW) under Contract No. 530-4CDP03; Ministry of Development of Turkey under Contract No. DPT2006K-120470; National Science and Technology fund; The Knut and Alice Wallenberg Foundation (Sweden) under Contract No. 2016.0157; The Royal Society, UK under Contract No. DH160214; The Swedish Research Council; U.S. Department of Energy under Contracts No. DE-FG02-05ER41374, No. DE-SC-0010118, and No. DE-SC-0012069; and University of Groningen (RuG) and the Helmholtzzentrum f\"ur Schwerionenforschung GmbH (GSI), Darmstadt.

% -- Bibliography --

\end{document}